\documentstyle[a4wide,12pt,epsf]{article}
\begin{document}
\begin{flushright}
\baselineskip=12pt
{RHCPP99-20T}\\
{astro-ph/9911485}\\
{November 1999}
\end{flushright}

\begin{center}
{\Large \bf 
A possible explanation of Galactic Velocity Rotation Curves in terms of a 
Cosmological Constant \\}
\vglue 0.35 cm
{Steven B Whitehouse 
\footnote {Sbwphysics@aol.com} and  George V. Kraniotis
 \footnote
 {G.Kraniotis@rhbnc.ac.uk}\\}
\vglue 0.35 cm
{\it  Centre for Particle 
Physics ,\\} 
{\it Royal Holloway College, University of London,\\}
{\it Egham, Surrey TW20-0EX, U.K.\\}


\baselineskip=12pt

\vglue 0.45cm
{\bf Abstract}
\end{center}
{\rightskip=3pc
\leftskip=3pc
\noindent
\baselineskip=20pt

This paper describes how the non-gravitational contribution to 
Galactic Velocity Rotation Curves can be explained in terms of 
a negative Cosmological Constant 
($\Lambda$).

It will be shown that the Cosmological Constant leads to a velocity 
contribution proportional to the radii, at large radii, and depending on 
the mass of the galaxy. This explanation contrasts with the usual 
interpretation that this effect is due to Dark Matter halos.

The velocity rotation curve for the galaxy NGC 3198  will 
be analysed in detail, while several other galaxies will 
be studied superficially.

The Cosmological Constant derived experimentally from the NGC 3198 data 
was found
 to be:
$|\Lambda|_{Exp}=5.0\times 10^{-56} cm^{-2}$.
This represents the lowest experimental value obtained from the 
set of galaxies studied and compares favourably with the 
theoretical value obtained from the Large Number Hypothesis of:
$|\Lambda|_{Theory}=2.1\times 10^{-56}cm^{-2}$.

It will be 
shown that the Cosmological 
Constant, in the Weak Field Approximation, leads to a 
correction term for the Newtonian potential and the 
corresponding acceleration of a test particle.
The Cosmological term leads to a modified Newton equation given by:
$F_{m_1}=m_1\Bigl[-\frac{G m_0}{r^2}+G_{\Lambda}r\Bigr].$

Here the force experienced by a mass $m_1$ is given by the 
sum of the gravitational fields produced by $m_0$ and $\Lambda$.
For a non-zero Cosmological Constant the modification force represents 
a fifth fundamental force, gravitational in nature, and proportional to 
the distance.

The Extended LNH is  then  used to define other cosmological parameters: 
gravitational modification constant, energy density, and 
the Cosmological Constant in terms of a fundamental length.

A speculative theory for the evolution of the Universe is outlined where 
it is shown how the Universe can be defined, in any 
particular era, by two parameters: the fundamental length and the energy 
density of the vacuum for that epoch (GUT, electroweak, Quark - Hadron 
Confinement).

The theory is applied to the time evolution of the universe where a possible
 explanation for the
$\rho_{Planck}/\rho_{\Lambda}^{QH} \approx 10^{120}$
problem is proposed.

Finally, the nature of the ''vacuum'' is reviewed along with a speculative 
approach for calculating the Cosmological Constant via formal M-theory.
The experimentally and theoretically derived results presented in 
this paper support a decelerating Universe, in contrast with recent 
results obtained from Type 1a Supernovae experiments, suggesting an 
accelerating Universe.

}

\vfill\eject
\setcounter{page}{1}
\pagestyle{plain}
\baselineskip=14pt

\section{Cosmological Constant, Dark Matter and Velocity Rotation Curves}

\subsection{ Short History of the Cosmological Constant}

	There has been considerable historical interest in 
the Cosmological Constant from the time Einstein added it to 
his Gravitational Field Equations. His aim 
was to artificially induce a 
stationary solution in order to support a static 
unchanging universe. In doing this he neglected to predict the 
Expanding Universe, a natural consequence of his theory, which was later 
developed by Hubble from astronomical observation.

In every decade since its inception the Cosmological Constant has been used 
to support the prevailing theory of the time. Particle physicists would like
 it to be identically zero in order to support the Standard Model, 
while cosmologists have taken negative, zero or 
positive values in order to predict either a 
contracting, expanding or accelerating universe.

Two excellent pedagogical reviews of the 
Cosmological Constant have been written by Cohn \cite{COHN}  
and Carroll $\&$ Press \cite{CARROLL}, while Turner 
\cite{turner} has reviewed the cosmological 
parameters in light of the latest experimental data.

Recent analysis \cite{SUPERN} of type Ia supernovae has 
led to the prediction of an Accelerating Universe. The study 
and fundamental understanding of the Cosmological Constant 
has again become a $\it{cause \;\;c\acute{e}l
\grave{e}bre}$.  It lies at the epicentre of several lines of 
fundamental research, namely: Astronomy,
Cosmology, String Theory, Inflation Theory, High Energy and Particle Physics.

If the Universe is accelerating or decelerating, it suggests that it is being 
driven apart 
or forced together by an exotic new form of energy.
A non zero Cosmological Constant, representing in 
some way the energy of the vacuum, would 
produce such a  force. 
The sign of the Cosmological Constant  will determine if the 
force acts as a 
kind of antigravity leading to an accelerating 
universe, or a new type of gravitational 
force, supplementing Newtonian gravity, resulting in a 
decelerating Universe. In either case it would be expected that there 
would be a deviation from the inverse square law.

\subsection{``Dark Matter'' - Energy Deficit of the Universe}
In 1844, Friedrich Bessel was the first to infer the existence of 
non-luminous Dark Matter from gravitational effects on positional 
measurements of Sirius and Procyon. He inferred that 
each was in orbit with a mass comparable to 
its own (See Trimble \cite{TRIMBLE} for a pedagogical review of Dark Matter).

In 1933, Zwicky concluded that the velocity dispersion in 
Rich Clusters of galaxies required 10 to 100 times 
more mass to keep them bound than could be 
accounted for by luminous galaxies themselves.

The old problem of "missing" matter is today 
referred to as the Energy Deficit of the Universe. The 
present consensus is that the energy of the universe
\cite{turner}, assuming a present day flat Universe ($k=0$), is 
comprised of two key
 components: matter, Baryonic and Non-Baryonic, and 
Dark Energy which is associated with 
the energy of the vacuum represented by the Cosmological Constant.

Baryonic and Non-Baryonic Matter represent approximately, $5\%$ and $35\%$ 
of the
 critical mass of the Universe while the 
Cosmological Constant (vacuum energy) accounts for $60\%$.

The need for Non-Baryonic Matter arises mainly becauseBaryonic
 Matter is limited to approximately $5\%$ of the Critical Mass of the Universe.
 This limitation is set by Big Bang Nuclear 
Synthesis considerations \cite{turner}.

There are many candidates for the Non-Baryonic components of matter: 
two presently of interest are \cite{DAVID}: WIMPs, Weakly Interacting 
Massive Particles, and Axions - postulated to 
explain the lack of CP violation in the strong interaction.

\subsection{Velocity Rotation Curves}

The experimental determination of Galactic Velocity Rotation Curves has been
 one of the mechanisms to estimate the "local" mass (energy) density of the 
Universe. In a paper reviewing Dark Matter Trimble 
\cite{TRIMBLE} noted that many Velocity Rotation Curves 
remained flat or even rise to large radii, well outside
 the radius of the luminous astronomical object. This non-Newtonian behaviour 
has lead to the developement of ``Dark Matter'' theories in order to explain 
the missing mass of the Universe.

The majority of the Velocity Rotation Curves are described, at large radii, 
in terms of some kind of "dark matter" component, results are often 
described in terms of "Dark Matter" halos (hollow halos) dominating 
the gravitational potential at some 
nominally large distance. However none of the present theoretical models 
successfully explain the observational data \cite{ITALY,SANCI}.

In this paper several Galactic Velocity Rotational 
Curves data sets will 
be analysed and the results presented in a later section.
The case will be made that the Cosmological Constant 
represents a constant energy term which contributes to 
Galactic Velocity Rotation Curves.

\newpage
\section{Experimental Results for Galactic Velocity Rotation Curves}
\subsection{ Approach}

This section will discuss and analyse in detail the velocity rotation curve 
data for galaxy  NGC 3198 
\cite{BAHCALL}. The galaxy has been studied extensively 
resulting in several sets 
of  experimental data that are suitable  
for 
testing theoretical predictions. Data for 
other randomly selected Galaxies: NGC 2403, NGC 
4258, NGC 5033, NGC 5055, NGC 2903, NGC 6503, will also be analysed in 
order to establish a general experimental trend.

The theory presented in the next section of this paper predicts that 
velocity rotation curves have 
two main components: gravitational mass and effective
 mass due to the Cosmological Constant. Following this approach, the 
gravitational contribution to the velocity 
rotation curve will be subtracted from the observational data. The 
resulting graphs are predicted to be straight lines whose 
gradients are proportional to the Cosmological Constant at large radii.

\subsection{ NGC 3198 Galaxy}

NGC 3198 is found in the constellation of Ursa Major and is 
classified as a Sc spiral galaxy. The galaxy has little or no 
prominent bulge and has a normal exponential disk. The observed 
rotation curve rises gradually near the centre and flattens out at 
approximately 150 Km/s in the outer region.

It has been exhaustively studied, van Albada et.al., 
van Albada $\&$ Sancisi, Kent, Begeman, \cite{BAHCALL,KENT} and others. For 
illustrative purposes the extended velocity rotation curve 
of NGC 3198, shown in figure 1, is taken 
from Albada et.al. (figure 7). It extends to eleven scales 
lengths which corresponds to approximately 50 Kpc.

The high quality data has allowed detailed theoretical 
analysis of the rotation curve. The digitised data, shown in 
Figure 1, was used to determine the gravitation and 
Cosmological Constant contributions to the velocity rotation curves.

Drawing a straight line through the data points in Figure 2 
yields a value of:
\begin{equation}
|\Lambda_{Exp}^{NGC 3198}|=5.0\times 10^{-56} cm^{-2},
\end{equation}
for the Cosmological Constant.

\begin{figure}[t]
\epsfxsize=6.9in
\epsfysize=8.1in
\epsffile{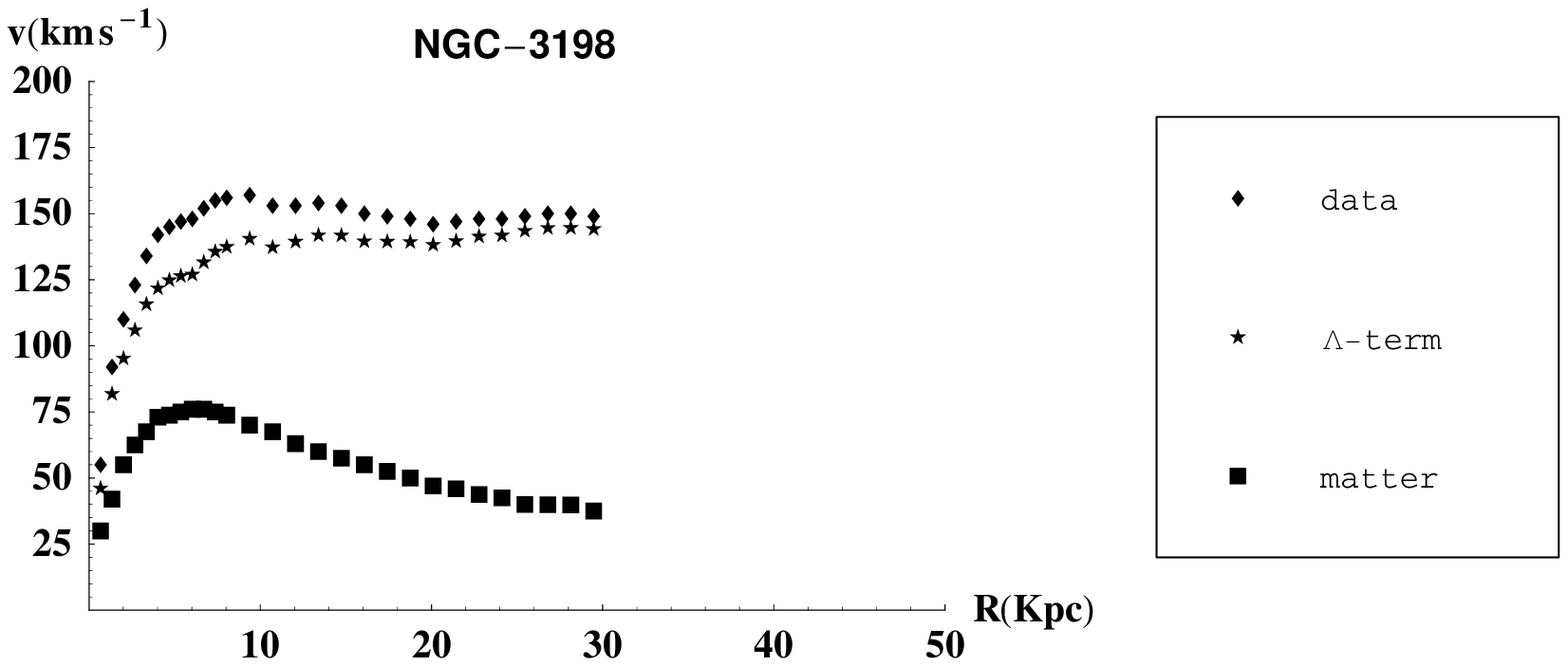}
\caption{Reproduced with the permission of Corbelli $\&$ Salucci.}
\end{figure}

\begin{figure}
\epsfxsize=6.9in
\epsfysize=7.9in
\epsffile{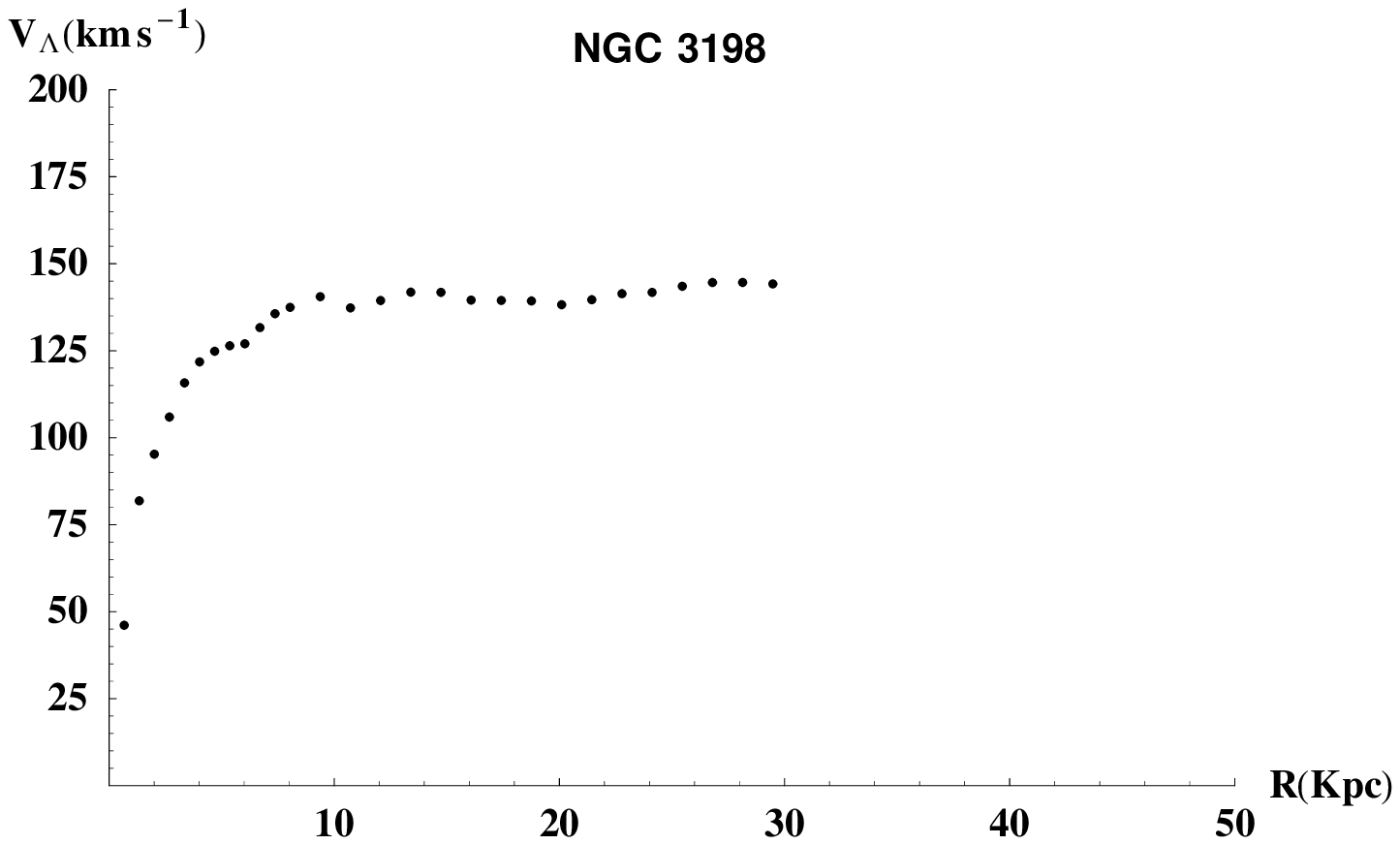}
\caption{Cosmological Constant contribution to the  NGC-3198 rotation curve}
\end{figure}


\subsection{ NGC 2403,3198,157,2841 Galaxies}

The velocity rotation curve data for galaxies, chosen at random: NGC 2403 
\cite{VAN}, NGC 3198 \cite{VAN2}, NGC 157
\cite{ZASOV}, NGC 2841 \cite{SANDERS}, were subjected to a 
crude gravitational and Cosmological Constant 
decomposition analysis. The gravitational contribution 
was subtracted from the full 
rotation curve data, and in all cases good  
straight lines were obtained for the Cosmological Constant 
contribution.

For the galaxies studied , the radii of the galaxies and the value 
for the Cosmological Constant and the associated 
densities, are shown in table 1.
\begin{table}
\begin{center}
\begin{tabular}{|c|c|c|c|}\hline\hline
${\rm Galaxy}$ & ${\rm Radius}$ & ${\rm Value\; of\; Cosmological\; Constant}$ & ${\rm Galactic\; Density}$
\\ \hline\hline

NGC 2403 & 20Kpc & $|\Lambda|_{NGC2403}=3.6\times 10^{-55} cm^{-2}$ &
$\rho_{\Lambda}^{Exp}=3.9\times 10^{-28} g\;cm^{-3}$ \\ \hline
NGC 4258 & 50 Kpc &$|\Lambda|_{NGC 4258}=5.5\times 10^{-55} cm^{-2}$ &
$\rho_{\Lambda}^{Exp}=5.9\times 10^{-28} g\;cm^{-3}$ \\ \hline
NGC 3198 &50Kpc & $|\Lambda|_{NGC3198}=5.0\times 10^{-56} cm^{-2}$ & 
$\rho_{\Lambda}^{Exp}=5.4\times 10^{-29} g\;cm^{-3}$ \\ \hline
NGC 5033 & 40 Kpc & $|\Lambda|_{NGC 5033}=1.0\times 10^{-55} cm^{-2}$ &
$\rho_{\Lambda}^{Exp}=1.1\times 10^{-28} g\;cm^{-3}$ \\ \hline
NGC 5055 & 50 Kpc & $|\Lambda|_{NGC5055}=1.4 \times 10^{-55} cm^{-2}$ &
$\rho_{\Lambda}^{Exp}=1.5\times 10^{-28} g\;cm^{-3}$ \\ \hline
NGC 2903 &30Kpc & $|\Lambda|_{NGC 2903}=3.8\times 10^{-55} cm^{-2}$ &
$\rho_{\Lambda}^{Exp}=4.1\times 10^{-28} g\;cm^{-3}$ \\ \hline
NGC 6503 &30 Kpc &  $|\Lambda|_{NGC 6503}=4.6 \times 10^{-55} cm^{-2}$ &
$\rho_{\Lambda}^{Exp}=4.9\times 10^{-28} g\;cm^{-3}$ \\
\hline\hline
\end{tabular}
\end{center}
\caption{The equations for all the Cosmological parameters 
are given in the next section.}
\end{table}

It is quite 
striking that for such a relatively crude analysis
the experimental values for the Cosmological Constant are so consistent
and lie within the allowed cosmological constant range \cite{GVKSBW}.

\subsection{Review of Experimental results and Comparison to Theory}
\subsubsection{Review of Experimental Data}

The experimental difficulties in obtaining galactic rotation curve data are 
acute \cite{TRIMBLE}. Apart from the instrumental use and 
calibration difficulties, assumptions 
have to be made concerning the different contributions to the 
galactic baryonic mass from the stellar disk, warm 
ionised gasses, atomic and molecular hydrogen.

These types of problems are well considered in 
van Albada et. al. \cite{BAHCALL}
study of NGC 3198. Several types of mass models were used in 
conjunction with the experimental data in order to obtain 
the "Dark Matter" component. Figure 7 from their paper 
leads to a value of the Cosmological Constant of:
$$|\Lambda|_{NGC 3198}=5.0\times 10^{-56} cm^{-2},$$
while figure 4, with a greater model mass contribution gives:
$$|\Lambda|_{NGC 3198}=1.0\times 10^{-55} cm^{-2}.$$

The analysis detailed in this paper applies to velocity rotation 
curves that extend to many radii, at which stage the 
rotation velocity is constant or slightly rising.

\subsubsection{Comparison with Theoretical Predictions}

The gradient determined from the experimental 
galactic velocity rotation curve data is the only information 
needed to determine the Cosmological Constant for the 
present day Universe. It is remarkable that for such a crude 
analysis the experimental results  are so consistent.

A representative experimental value for the 
Cosmological Constant taken from table 1 is 
$|\Lambda_{Exp}|\sim 2.0\times 10^{-55} cm^{-2}$. This compares 
favourably to the theoretical value predicted by the Extended LNH of  
$|\Lambda_{Theory}^{QH}|=2.1\times 10^{-56} cm^{-2}$.

Again, the agreement between experimental and theory is striking. The 
factor of 10 between the two results maybe significant. However, it 
will not be possible to comment further until a 
more fundamental and systematic analysis of galactic 
velocity rotation curve data has been completed.

Finally, an effective mass density contribution to the Universe 
due to the Cosmological Constant can be determined (see table 1 above). The 
values given for experiment and theory are:  
$\rho_{\Lambda}^{Exp}\sim 4 \times 10^{-28} g\;cm^{-3}$ and 
$\rho_{\Lambda}^{Theory}\sim 1.2 \times 10^{-29} g\;cm^{-3}$ 
respectively. These can be compared with
 the best estimates for galactic Baryonic mass density 
which are in the range \cite{KOLB,GRAVITATION}:
$$10^{-29}>\rho_B>2\times 10^{-31} g cm^{-3}.$$

\subsubsection{ Negative Cosmological Constant}

Recently there has been great interest in the Type Ia Supernovae results 
of Perlmutter et al \cite{SUPERN} which suggest that the universe 
is accelerating.\\
\\
In this section we will show that the Weak Field Approximation 
coupled with galactic velocity rotation curve data inevitably lead to a 
negative Cosmological Constant.\\

The equation for the VRC is given by
\begin{equation}
-\frac{v^2}{r}\approx -\frac{G m}{r^2}+\frac{c^2 \Lambda}{3} r
\label{improv}
\end{equation}
(this expression 
is derived in section 3.3). We note that Eq.(\ref{improv}) is 
only strictly true for small and 
large radii, however it will serve to illustrate our 
arguments.\\

At small galactic radii the velocity versus radius contribution is 
well known and follows Newtonian physics. For large radii, a negative 
Cosmological Constant gives  a positive contribution to the 
VRC, which is what is actually observed.
On the other hand, the effect of a positive Cosmological Constant  
would be  to lower the rotation curve below that due 
to matter alone. \\
\\
The above simple argument, based on observational astronomy, 
allows only a negative Cosmological Constant as a possible explanation  
for the galactic velocity rotation curve data.
This is in clear disagreement with the Type Ia supernovae results 
\cite{SUPERN}. However, given  the uncertainties in the determination 
of the deceleration parameter, $q_0$, derived from supernovae data 
\cite{SUPERN} the conclusions  
outlined  above have certain merits worth consideration. \\
\\
In summary these are:  the 
Cosmological Constant is determined from $direct$ measurement 
unlike the Supernovae results, the experimentally determined value 
is  the correct 
order of magnitude as that required from cosmological constraints, and 
lastly a negative Cosmological Constant is consistent, and indeed 
a natural physical explanation, for the observed galactic 
velocity rotation 
curve data.\\
\\
Finally, observations of global clusters of stars constrain 
the age of the universe and consequently place an observational limit 
on a negative Cosmological Constant \cite{OHANIAN} of: 
\begin{equation}
|\Lambda| \leq 2.2 \times 10^{-56} cm^{-2}.
\label{limit}
\end{equation}

This upper limit value for the Cosmological Constant derived from 
global cluster constraints is in agreement with the 
experimentally determined value derived from galactic velocity 
rotation curve data.

\subsection{ Review of Experimental Results and Energy Deficit of the Universe}

A systematic and general study of galactic velocity rotation curves needs to
 be undertaken in order to confirm the results presented in this paper, and 
to establish that there are no other ("Dark Matter") components that 
contribute to the velocity rotation curves.

Superficial results (NGC 3198) suggest that 
the Cosmological Constant could be a 
candidate for the "dark energy" which is smoothly distributed and contributes 
approximately $60\%$ to the critical density of the Universe \cite{turner}. 
This assertion is supported by the successful 
application of the "Cold-Dark Matter" model [7, 14, 15] in predicting 
mass and galactic structures formation.

The ranges in the values for the density contributions of the Cosmological 
Constant and Baryonic matter, make it difficult to 
comment further on 
the larger problem of the Energy Deficit of the Universe.

The cosmological parameters are very sensitive to small changes in 
the values of the String length and the 
experimentally derived Cosmological Constant. This sensitivity in 
conjunction 
with the rapidly improving accuracy of astronomically 
derived data [9] will combine to provide a powerful tool 
for probing the Universe.

Finally, the experimental results 
presented in this paper predict a fifth force, gravitational in nature, 
proportional to the distance between bodies, leading to a 
decelerating Universe.
\newpage
\section{ Theoretical Framework}
\subsection{ Extended Large Number Hypothesis, 
Duality, Weak Field Approximation}

The starting point for the work presented in this paper was a firm belief in
 two key points: that nature has a fundamental length and that the 
Cosmological Constant represents in some way a 
macroscopic quantum mechanical parameter.

These guiding objectives will be explored using String Theory, the Extended 
Large Number Hypothesis and the Weak Field Approximation. Duality 
arising out of String Theory 
points to a fundamental length, the Extended Large Number
Hypothesis show how macroscopic Cosmological Parameters can be related to 
quantum mechanical origins, and 
facilitates the estimation of their values, and 
finally the Weak Field Approximation allows the 
Cosmological Constant to be defined in terms of an effective mass.

\subsubsection{ Large Number and Extended Large Number Hypothesis}

Extending the work of Eddington \cite{ARTHUR} Dirac wrote a 
paper titled, ``A new basis
 for Cosmology'' \cite{paul}, where he considered 
the strange coincidence between the ratio of 
Cosmological and Atomic Constants to the 
approximate age of the Universe, namely:
\begin{equation}
\frac{hc}{G m_p m_e}\sim \frac{m_p c^2}{H h}\sim 10^{41},
\label{LANUMBER}
\end{equation}
Here the masses represent that of the proton and electron and $H$ is Hubble's 
Constant.

Dirac's proposal became known as the Large Number Hypothesis (LNH) \cite{
paul,DIRAC,LARGE}.
 He firmly believed that these ``coincidences'' represented  an as 
yet unknown fundamental theory linking the 
Quantum Mechanical origin of the Universe to the 
large scale cosmological parameters.

Zel'dovich \cite{ZELDO}, in an 
extraordinary paper, extended the LNH to include an expression 
for the Cosmological Constant, given by
\begin{equation}
|\Lambda|=\frac{8 \pi G^2 m_p^6}{h^4}
\label{COSMO}
\end{equation}
 He went on to show that the Cosmological Constant in empty space produces 
the same gravitation field as when the space contains matter, and that these
 terms enter as fully fledged terms, in the presence of ordinary matter, in 
the Gravitational Field Equations.

Starting from a theory for elementary particles, Zel'dovich then showed that 
the vacuum energy for non-interacting particles was identically equal to 
zero. However when the interactions between particles were taken 
into account, it resulted in a  non-zero value for the vacuum energy.

Zel'dovich interpreted the gravitational energy of the vacuum in terms of 
the interaction of virtual particles, the distance between them being defined 
by $\lambda=h/m_p c$. Here the self energy is exactly 
equal to zero, but the 
gravitational interacting of neighbouring particles caused 
the vacuum energy to have a non-zero value, given by:
\begin{equation}
\epsilon_{vac}=\frac{G m_p^6 c^4}{h^4}
\label{density}
\end{equation}
Sakharov \cite{ANDRE} further extended LNH by 
proposing  a gravitational theory or 
justification of the equations of General Relativity theory 
based on the considerations of vacuum fluctuations. He stressed the 
importance of the hypothesis that there is a fundamental 
length or corresponding limiting momentum, less than which 
the theory is not valid. He proposed this limit as :
$k_{max}\sim 10^{33} cm^{-1}$, leading to:
\begin{equation}
L^2=\frac{G h}{c^3} 
\label{FLENGTH}
\end{equation}
Zel'dovich concluded his paper by proposing that the Cosmological Constant 
could be interpreted as follows:

 ``there exists a theory of elementary particles which would give (in 
accordance with a mechanism which is still unknown at present) an 
identically vanishing energy, provided that this theory 
was applicable without limit, up to arbitrary large momenta; 
there exists a momentum $p_{max}$, beyond which the theory is
 invalid.''

This theory is interpreted as the present day String Theory.

Finally, Matthews demonstrated in an inspirational article \cite{MATTHEWS}, 
that the mass, defined in terms of the mass of a Proton, 
introduced by Dirac in an arbitrary way, should be replaced by 
the effective (quantum) mass associated with the various phase 
transition energies of the Universe (GUT, electroweak, Quark-Hadron).

He showed that using the value for the vacuum energy for the 
last phase transition, Quark-Hadron, leads to the present 
observed cosmological parameters, and that 
current cosmic dynamics were seemingly determined by the aftermath of 
the most recent quantum vacuum state transition.

He also pointed out that the energy density required to 
support inflation was that associated with the GUT phase transition.

\subsubsection{ Fundamental Length and Duality}
$ 3.1.2.1 \;Duality - Fundamental\; Length\; Justification$

Zel'dovich and Sakharov understood the need for a theory of elementary 
particles that inevitably leads to a 
concept of a fundamental length. The closest, and indeed, the 
only theory that is a candidate today is String theory,  
it provides a  consistent framework in which one 
can study quantum gravity and its unification with the 
three gauge forces \cite{STRING}.

One of the most profound quantum symmetries of String theory, 
with no field theory analogy, is (target space) Duality 
(for a review see Giveon et al \cite{GIVEON}).  
In its simplest form, Duality shows that a (closed) string 
moving on a circle of radius $R$, is equivalent 
to one that moves on a circle of radius  $R\rightarrow 
\alpha^{'}/R$ when the 
momentum modes ($n/R$) are interchanged with winding modes 
$mR/\alpha^{'}$ and $\alpha^{'} \simeq (10^{-32} cm)^2 $ controls 
the tension of the string. The above symmetry can 
be extended to more complicated situations 
when background fields are present in six-dimensional compactifications 
of string on orbifolds or Calabi-Yau manifolds. There are some 
interesting  consequences 
of  Duality. One is the prediction of a 
fundamental minimal observable length scale of order   
$\sqrt{\alpha^{'}}$ in Nature that leads to the natural 
generalisation of the Heisenberg Uncertainty and 
Equivalence principles \cite{VENE}. Another  
is that it  restricts the possible form of scalar 
(or super) potentials and determines some 
characteristics of non-perturbative supersymmetry-breaking and 
CP violation \cite{SCALAR,DBGVKAL}.

Duality indicates that Einstein's 
action will be modified drastically
 at short distances, and the approach 
offered by Duality may help to resolve problems 
associated with Einstein's Cosmology, including the 
initial singularity problem.
\\

The generalised Uncertainty Principle becomes:
\begin{equation}
\Delta x \geq \frac{\hbar}{\Delta p}+\alpha^{'} \frac{\Delta p}{\hbar }
\label{Uncer1}
\end{equation}
where 
\begin{equation} 
\Delta x \geq L_{Planck}=\sqrt{\frac{h G}{c^3}}.
\label{HEISE}
\end{equation}
For a recent discussion on the theoretical attempts to derive the second term 
in equation (\ref{Uncer1}) we 
refer the reader to \cite{NICK}.
The need for a fundamental length was confirmed 
by Veneziano et al \cite{VENE} who,
 when studying fixed angle high energy scattering processes, found that 
strings were not applicable 
to distances shorter than the string length. This effect 
was also confirmed by Gross $\&$ Mende \cite{GROSS} 
who studied the high-energy behaviour 
of string scattering amplitudes to all order in perturbation theory.


In conclusion, the fundamental length 
implied by Duality is 
related beautifully by Sakharov and Zel'dvich in their proposal 
where they  envisioned a theory of fundamental 
length before the advent of String Theory.

\subsection{ Relationship between Quantum Mechanics and General Relativity}

One of the fundamental aims in the 20th Century has been the formulation of 
a theory relating Quantum Mechanics, a possible 
candidate for the creation of the Universe, to General Relativity, 
the theory of physics for macroscopic masses and distances.

The approach taken by Zel'dovich \cite{ZELDO} was to 
suggest that gravitational interactions of 
virtual particles in vacuum would endow empty space with an 
effective energy and pressure, which he 
interpreted to be the Cosmological Constant term, $\Lambda$. 
This led Sakharov \cite{ANDRE} to suggest that there should be 
a functional relationship between the 
quantum nature of the vacuum and the curvature of space. He 
introduced a momentum cut off of $k_0$ for the virtual particles, 
which he stated should be given, 
{\it a priori}, by a microscopic field theory. 
Target-space duality invariant, String Theory, is 
the microscopic theory that leads to a 
fundamental length scale or momentum $k_0$, which in turn  determines the 
curvature of space-time and the Cosmological Constant. The 
Planck compactification radii are preferred by 
Duality, and moreover, coincide with the 
value of $k_0\sim 10^{33} cm^{-1}$ , which 
gives the correct Newtonian gravitational constant. This 
result is regarded as an important evidence for the relevance of Duality in Nature.

\subsection{Effective Mass due to Cosmological 
Constant within the Weak Field Approximation}
\subsubsection{ Effective mass due to the 
presence of the Cosmological Constant}

This section derives  the contribution to  
the galactic  rotation due to 
the Cosmological 
Constant within the Weak Field Approximation.

Following Zel'dovich, the Einstein Field equations can be rewritten as:
\begin{equation}
R_{\alpha\beta}-\frac{1}{2}g_{\alpha\beta}R=
\frac{8\pi G}{c^4}\Biggl(T_{\alpha\beta}+
\frac{c^4\Lambda}{8 \pi G}g_{\alpha\beta}\Biggr).
\end{equation}
The assumption, $\Lambda \not = 0$ , 
denotes that empty space produces the same gravitational field as 
when the space contains matter defined as:
\begin{equation}
\rho_{\Lambda}=\frac{c^2 \Lambda}{8\pi G};\;\epsilon_{
\Lambda}=\frac{c^4 \Lambda}{8\pi G};\;P_{\Lambda}=-\epsilon_{
\Lambda},
\label{VACUUM}
\end{equation}
where $\rho_{\Lambda}, \epsilon_{
\Lambda}, P_{\Lambda}$
represents the effective mass density, energy density and 
pressure due to the presence of the Cosmological Constant. Here 
the energy density and pressure have been 
formulated in such as way as to leave the theory relativistically invariant.

Following the approach of Ohanian $\&$ Ruffini 
\cite{OHANIAN}, the Weak Field 
approximation to the Field Equations gives rise, 
in the absence of matter, to a differential 
equation for the Newtonian potential described as:
\begin{equation}
\nabla^2 \Phi=-c^2 \Lambda.
\label{DYNAMIKO}
\end{equation}
Comparing this with Newton's equation:
\begin{equation}
\nabla^2 \Phi=4\pi G \rho,
\label{NEWTON}
\end{equation}
the $ \Lambda$ is recognised  
to correspond to a uniform effective mass density given by 

\begin{equation}
\rho_{\Lambda}^{eff}=-\frac{c^2 \Lambda}{4 \pi G}.
\end{equation}

It is important to realize that 
the Cosmological Constant obeys the equation of state given by:  
\begin{equation}
P_{\Lambda}=-c^2 \rho_{\Lambda},
\label{state}
\end{equation}
Taking the Newtonian limit in the 
absence of matter, $T_{\mu\nu}=0$, the differential equation for 
the static 
Newtonian potential becomes
\begin{equation} 
\nabla^2 \Phi=-c^2 \Lambda
\label{potential}
\end{equation}
leading to: 
\begin{equation}
\rho_{eff}=\rho_{\Lambda}+\frac{3 P_{\Lambda}}{c^2}=-2\rho_{\Lambda}.
\end{equation}
Therefore the factor of two and the change in sign is due to the fact 
that the Cosmological Constant obeys (\ref{state}) and that in the 
linearised theory the source term for 
$h_{00}$ is $T_{00}-\frac{1}{2}Tr T_{\mu\nu}$ \cite{OHANIAN}.

By arbitrarily setting $\Phi=0$ at the origin, and 
using spherical polar coordinates, the solution for $\Phi$ becomes:
\begin{equation}
\Phi=-\frac{\Lambda c^2 r^2}{6}
\end{equation}
The potential indicates that between any two particles for 
$\Lambda>0$ and $\Lambda<0$, there acts a repulsive force 
or attractive force which is proportional to $r$. This force represents a 
new fundamental force, a fifth force, which is gravitational in nature.

In the presence of matter and using the Weak Field 
Approximation the modified Newtonian potential is given by:
\begin{equation}
\phi_{MN}=-\Biggl[\frac{G m_0}{r}+G_{\Lambda}r^2\Biggr].
\label{WEAKFIELD}
\end{equation}
The acceleration experienced by a test particle following a geodesic 
path created by the gravitational fields produced by 
$m_0$ and $\Lambda$, is given by:
\begin{equation}
\ddot{x}=-\frac{G m_0}{r^2}+G_{\Lambda}r
\end{equation}
or when applied to galactic rotation curves becomes:
\begin{equation}
-\frac{v^2}{r} \approx  \frac{-G m_0}{r^2}+G_{\Lambda}r,
\label{APTRUE}
\end{equation}
where $m_0$ represents the mass of the galaxy, and $G$ and 
$G_{\Lambda}$ are the gravitational and anti-gravitational constants.

The above equation is an approximation that is only strictly correct 
for small and large radii. In addition to this the second term, arising 
from the 
Weak Field Approximation, is valid only when the effective density 
due to $\Lambda$ is much greater than the gravitational density. 
This condition is only usually satisfied at many galactic radii.

The two terms should not be equated in order to provide estimates for 
a critical radii where effects due to Newtonian gravity and the 
Cosmological Constant are in  balance as proposed by Bergstrom \cite{
SWEDEN,SWEDEN2} 
following the first version of this paper.

This procedure leads to the incorrect conclusion that effects due to 
the Cosmological Constant are not significant on galactic scales.  To use 
the approach of Bergstrom a Strong Field Approximation would 
have to be developed.

[Note: The equation for the acceleration experience by a body is only 
strictly true for a test particle. The mass of 
the body itself produces curvature 
that affects the original geodesic.  The gravitational 
interaction of two massive bodies is not 
directly addressed by Einstein's theory of 
General Relativity; however approximate 
methods were developed in the 
1980's ( Damour $\&$ Deruelle, 1986)\cite{DAMOUR}).

\subsubsection{ Valid Ranges for Weak Field Approximation}

The Weak Field approximation is valid for the following range of values 
\cite{OHANIAN}:
\begin{equation}
\sqrt{\frac{1}{\Lambda}}\gg r \gg \frac{G M}{c^2}.
\end{equation}

This approximation is strictly valid for astronomical masses and 
distances in the range  below:
\begin{equation}
\sqrt{\frac{2}{\Lambda_{Exp}}}\sim 1447.3 Mpc,\;\;
\sqrt{\frac{2}{\Lambda_{Theory}}}\sim 2233.2 Mpc,\;\;\frac{G M}{c^2}
\sim 47.7 pc
\end{equation}
Here the mass of a Supercluster \cite{COSMO} has 
been used, along with the experimental and theoretical 
derived values for the Cosmological Constant. The approximation 
defined above is found to be valid for all astronomical objects of 
interest.

\subsection{ Definition of Cosmological Constants in terms of a Fundamental 
Length}
Listed below are the expressions for the Cosmological 
Parameters that have been defined in terms of a 
fundamental length and the Cosmological Constant.
[Note: When defining the Cosmological Constant within the 
Entended LNH we have followed Dirac \cite{paul,DIRAC,LARGE} and  
Matthews \cite{MATTHEWS} in 
using $h$ instead of $\hbar$. Also in order to 
be consistent with general relativity we have 
used Kardashev's \cite{KARDA}  expression for the Cosmological Constant which 
includes a $8\pi$ multiplying term.]\\
\\
{\bf{Gravitational Constant}:}
\begin{equation}
G(L_s)=\frac{c^3 L_s^2}{h}.
\label{GCON}
\end{equation}

(Throughout this paper the experimental value of 
the Gravitational Constant will be used to define 
the fundamental length. The value is given by $L_s=1.6 \times 10^{-33} cm$).
\\

{\bf{Cosmological Constant}:}
\begin{equation}
|\Lambda(L_s)|=\frac{8 \pi c^6 m_{QH}^6}{h ^6}L_s^4,
\end{equation}
Here the Proton mass has been replaced by an effective mass (quantum 
energy scale) given by the Quark-Hadron phase transition vacuum energy.
\\

{\bf{Gravitational Modification Constant:}}
\begin{equation}
G_{\Lambda}(L_s)=\frac{c^2 \Lambda}{3}.
\end{equation}
Mass density (due to the presence of the Cosmological Constant):
\begin{equation}
\rho_{\Lambda}^{eff}(L_s)=\frac{-c^2 \Lambda}{4 \pi G}.
\end{equation}
\\

{\bf{Modified Newton Equation:}}
\\

Associating the effective mass density, $\rho_
{\Lambda}$ , with a cosmological modification force, 
Newton's equation can be modified and written as:
\begin{equation}
F_{m_1}(L_s)=m_1\Biggl[\frac{-G m_0}{r^2}+G_{\Lambda}r\Biggr],
\end{equation}
where  $G$ and $G_{\Lambda}$ are the 
Gravitational and  Gravitational Modification Constants respectively.

A positive value of $\Lambda$ leads to a repulsive or anti-gravitation force, 
whereas a negative value leads to an attractive and contracting force.
\newpage
\section{\Large{Speculative Theory for the evolution of the Universe}}
\subsection{ Evolutionary Universe}
In this section of the paper a speculative theory for the evolution of 
the Universe will be developed. It will build upon Dirac's belief that 
the present day cosmological parameters are 
historically connected to the quantum mechanical origin of 
the Universe; the connection being the Cosmological Constant, a 
macroscopic quantum mechanical parameter.

As the Universe evolves through the various quantum vacuum phase transitions
: GUT, Electro-Weak, Quark-Hadron, the value of the Cosmological 
Constant changes. The value is uniquely 
determined by the energy density of the vacuum for that epoch 
(epoch energy scale (EES)).

The relationship between the Cosmological Constant and the 
quantum mechanical effective mass is shown below:
\begin{equation}
|\Lambda_{EES}|=\frac{8 \pi c^6 m_{EES}^6}{h^6}L_s^4.
\end{equation}
(Note: The Cosmological Constant is not identical to the vacuum energy scale 
but rather to the gravitational interaction energy of the vacuum for that
 epoch. The distinction and clarification of these terms will 
be discussed later).

This equation relates the quantum mechanical origin of the Universe, in 
terms of an effective mass, to the macroscopic evolution 
of the Universe 
via the Cosmological Constant and Einstein's Field Equations.

Matthews \cite{MATTHEWS} inferred that the Universe's evolution was determined
 by the symmetry breaking vacuum phase transition. He suggested replacing 
the mass of the proton in the 
LNH by an effective mass determined by the Planck, GUT, 
Electro - Weak or Quark- Hadron energy scales.

\subsection{ The Two Parameter Universe}

Einstein had a philosophical belief in the simplicity of 
the laws of Nature and the Universe. This requirement for 
simplicity is supported by a theory where the Universe 
can be uniquely described by two parameters. These parameters, within 
the Extended LNH, are: a fundamental length and an equivalent effective 
mass  for the 
vacuum energy density for that epoch. 

The following sections review the various 
epochs of the Universe.

\subsection{The Present Epoch - Quark-Hadron Era}
This section will derive the values for the cosmological 
parameters for the present epoch. It will 
be assumed that the Universe to a first approximation is 
flat, the only curvature being that due to the Cosmological Constant.

The Universe that we see and observe today, defined by the present day 
cosmological parameters, is determined in 
great part by the energy density (scale) of the 
last phase transition - Quark-Hadron.

The values for the String length and effective mass of the 
Quark-Hadron phase transition \cite{MATTHEWS} are taken to be:
\begin{equation}
L_s=1.6\times 10^{-33} cm,\;m_{QH}\sim 0.15\;{\rm GeV}\equiv 2.5\times 10^{-25}g,
\end{equation}
which in turn lead to values for the Cosmological Constant 
and energy density of:
\begin{equation}
|\Lambda_{QH}^{Theory}|=\frac{8 \pi c^6 m^6_{QH}}{h^6}L_s^4=
2.1\times 10^{-56}cm^{-2},\;\rho_{\Lambda}^{Theory}=1.2\times 10^{-29}
g cm^{-3}.
\end{equation}

These values compare favourably to the experimental derived values of:
\\
$|\Lambda_{QH}^{Exp}|\sim 2.0 \times 10^{-55} cm^{-2}$, and $
\;\rho_{\Lambda}^{Exp}=4.0\times 10^{-28}g\; cm^{-3}$, respectively
and the generally accepted galactic Baryonic matter density which is 
in the range 
$1\times 10^{-29}\geq \rho_{Baryonic}\geq 2\times 10^{-31} g \;cm^{-3}$ 
\cite{GRAVITATION}.\\
\\
 The theoretical and experimental values for the 
cosmological parameters agree within a 
factor of ten. This agreement is good considering the 
magnitude of the parameters being used and the fact that the 
effective mass has been arbitrarily set to the vacuum energy scale 
of the epoch. Another explanation could be that there is 
another contribution to the missing "smooth dark
 energy" of the Universe other than the Cosmological Constant.

\subsection{ Planck, GUT, Electro-Weak and Quark-Hadron Epochs}
\subsubsection{Field $\&$ String Theories}
A non-zero vacuum energy presents present day Field $\&$ String Theories 
with a major problem, essentially the energies are too high to explain 
physical phenomena.

In general vacuum energies in quantum field theories are described by:
\begin{equation}
u_{vac}=\int_0^{E_{max}} \frac{d^{d-1}{\bf k}}{(2 \pi)^{d-1}}
\frac{\omega_{\bf k}}{2}
\end{equation} 
where $\omega_{\bf k}^2={\bf k.k}+m^2$. Here the vacuum energy density is defined as the sum of the 
zero-point energies of the modes of the field, and  $k$ represents the 
wave number.
The vacuum energy provides a source term in Einstein's Field Equations:
the Cosmological Constant, which should lead to observational effects.

In four dimensions we find the well known expression for the energy density,
\begin{equation}
u_{vac}\sim (E_{max})^4/(h c)^3 J/m^3
\end{equation}

Listed in table 2 below  are the equivalent mass 
densities for the various epochs that will be used for 
comparative purposes.

String theories, using here Bosonic String theory for illustrative reasons, 
can be described in terms of the following spacetime action \cite{POLCHI}:
\begin{equation}
S = (\frac{1}{2 k_0^2}) \int d^D x(-G)^{1/2} e^{-2\Phi}[-
\frac{2(D-26)}{3\alpha^{'}}+R \\ 
\\
\;\;-\frac{1}{12} H_{\mu\nu\lambda} H^{\mu\nu\lambda} 
 + 4 \partial_{\mu} \Phi \partial^{\mu} \Phi+0(\alpha')]
\end{equation}

Here, $\Phi$  represents the dilaton, $G_{\mu\nu}$ the metric of 
D dimensional space-time, 
$\alpha^{'}$
is related to the string tension, $H_{\mu\nu\lambda}$ the 
3 index field strength related to the anti-symmetric tensor 
$B_{\mu\nu}$ and $R$ the Ricci tensor.

The one-loop vacuum energy density in Bosonic  String theory is 
non-zero and of the order $(10^{18} GeV)^4$, since
$O(\alpha^{'})\sim 10^{18} GeV$ which corresponds to the String scale.

Even in softly broken supersymmetic string theories, it is expected that 
$\rho_{\Lambda}$ should be of the order
$\rho_{Electro-Weak}\sim 10^{23}g cm^{-3}$, which is approximately 
52 orders of magnitude larger that the present observable value.

As pointed out by Polchinski\cite{POLCHI}, "the Cosmological Constant is telling us 
that there is something we do not understand in Field and String theories 
about the vacuum. It constitutes one of the best clues for a unified =
theory with gravity !"

\begin{table}
\begin{center}
\begin{tabular}{|c|c|}  \hline\hline
{\bf Energy Scale} & {\bf Density} \\
\hline\hline
Planck$\;\;\sim 10^{19}$GeV  & $\rho_{Theory}^{Planck}\sim 0.93 \times 10^{91}
g cm^{-3}$ \\
\hline\hline
GUT$\;\;   \sim 10^{16}$ GeV & $\rho_{Theory}^{GUT}\sim 0.93 
\times 10^{78}g cm^{-3} $ \\
\hline\hline
Electro - Weak$\;\;\sim 300$ GeV & $\rho_{Theory}^{EW}\sim 0.93 \times 10^{22}g cm^{-3} $ \\
\hline\hline
Quark - Hadron$\;\;\sim 0.15$ GeV & $\rho_{Theory}^{QH}\sim 4.7 \times 10^{11}g cm^{-3}$ \\
\hline\hline
\end{tabular}
\end{center}
\caption{mass densities at various epochs using the QFT result.}
\end{table}

\subsubsection{Extended Large Number Hypothesis}
The value for the Cosmological Constant and the density will have 
varied depending on the particular 
evolutionary era of the Universe. These are shown in table 3 below  which
relates, energy scales, cosmological parameters 
and densities to the associated epochs.

\begin{table}
\begin{center}
\begin{tabular}{|c|c|c|}  \hline\hline
{\bf Energy Scale} & {\bf Cosmological Constant} & {\bf Density} \\
\hline\hline
Planck$\;\;\sim 10^{19}$GeV  & $\Lambda_{Theory}^{Planck}\sim 
1.9 \times 10^{63}cm^{-2}$ & $\rho_{Theory}^{Planck}\sim 1.0 \times 10^{90}
g cm^{-3}$ \\
\hline\hline
GUT$\;\;   \sim 10^{16}$ GeV & $\Lambda_{Theory}^{GUT}\sim 
1.9 \times 10^{45}cm^{-2}$ & $\rho_{Theory}^{GUT}\sim 1.0 
\times 10^{72}g cm^{-3} $ \\
\hline\hline
Electro - Weak$\;\;\sim 300$ GeV & $\Lambda_{Theory}^{EW} \sim 
1.4\times 10^{-36}cm^{-2}$ & $\rho_{Theory}^{EW}\sim 7.4 \times 10^{-10}g cm^{-3} $ \\
\hline\hline
Quark - Hadron$\;\;\sim 0.15$ GeV & $\Lambda_{Theory}^{QH} \sim 
2.1\times 10^{-56}cm^{-2}$ & $\rho_{Theory}^{QH}\sim 1.2 \times 10^{-29}g cm^{-3}$ \\
\hline\hline
\end{tabular}
\end{center}
\caption{Cosmological Constant and mass densities at various epochs in the 
ELNH model.}
\end{table}
The mystery of the 
$$\rho_{\Lambda_{Planck}}^{Planck}/\rho_{\Lambda_{QH}}^{QH} \approx 10^{120},$$
suggested by Weinberg \cite{STEVEN} as the 
``the worst failure of an order-of-magnitude estimate 
in the history of science'' , can now be  understood easily. 
The ratio is indeed correct, but clearly it compares  
energy densities for different epochs. (The value found 
from data taken from  table 3 gives 
$\rho_{\Lambda_{Planck}}^{Planck}/\rho_{\Lambda_{QH}}^{QH} 
\approx 8.8 \times10^{118}$).

The values of the cosmological parameters for the GUT scale 
are consistent with those required to drive inflation \cite{MATTHEWS}
but with a positive value for the Cosmological Constant. A possible 
mechanism to explain a changing sign for  
$\Lambda$ will be discussed later in the paper.
\subsubsection{Comparison of Field $\&$ String Theories with 
Extended Large Number Hypothesis}

It is also noted that while fundamental theories of Particle 
Physics such as the Standard Model, Quantum Field Theory and 
String Theory have many major predictive successes they all 
have problems with a high vacuum energy density.

On the other hand, while the Extended LNH is formulated from a 
naive theory \cite{ZELDO}, it appears to  
predict correctly the magnitude of the vacuum energy density and other 
cosmological parameters.

The ELNH seems to indicate that the gravitational self energy plays the 
key and central role in determining the magnitude of the vacuum energy. This 
could be an important clue towards the  theory of Quantum Gravity and 
suggests a starting point 
for a more formal approach.

\newpage
\section{ In the Search for a Theory of Quantum Gravity}
\subsection{The Vacuum}
The concept of a vacuum (state) has been invoked since the time of Faraday 
where it was used in relation to the ether. Over time it 
has come to mean different things to the various practitioners in the 
different fields of physics. 
\\

Questions concerning the vacuum could be:

\begin{itemize}
\item
What is the vacuum? How is it defined ?
\item
What is the connection between the ``real'' world and the vacuum ? 
Are they the same ?
\item
How do we determine the energy of the vacuum ?
\item
How does the vacuum ``communicate'' with the ``real world'' ?
\end{itemize}

These questions are of particular interest when considering a 
theory of gravity, and will be discussed in this section.

It is well known that the vacuum has certain properties: it represents the 
lowest energy state, it is Lorentz invariant and has a 
zero four-momentum. It may also carry 
quantum numbers like: isospin, parity and 
strangeness etc. \cite{LEE}, and is often 
considered as a ``medium'', analogous with a 
dielectric material, of  infinite  extent \cite{ZELDO}.

Ziman \cite{ZIMAN} defined a ``true'' vacuum 
state and then went on to show that not 
only is the state hypothetical, not physically accessible, but there 
is no mathematical description of it in terms of excited states (problem of 
infinities).

The vacuum or vacuum energy density is of paramount importance in cosmology 
and is often described in terms of scalar fields. Extreme lengths are taken 
to justify its inclusion. 

``... the advent of a constant homogenous scalar field, over all space 
simply represents the restructuring of the vacuum, in some sense, space filled 
with a constant scalar field does not carry a preferred reference frame 
with it, it does not disturb the motion of 
objects passing through space that it 
fills, and so forth. But when a scalar field appears, there is a 
change in the vacuum energy density... If there were 
no gravitational effects, this 
change in energy would go unnoticed \cite{ALIN}.''

At present, there is no experimental justification for the 
existence of scalar fields. However, from a theoretical standpoint, 
scalar fields play an important role, creating masses in 
the theory of elementary particle (Higgs mechanism), and 
scalar particles such as the Dilaton are always present in 
supersymmetric String theories.

For Cosmological reasons, among others, the vacuum energy density is 
associated with the energy scales of the Planck, GUT, 
Electro-Weak and Quark-Hadron
 eras. The question is: are these vacuum energy densities 
reflecting the ``real world'', and if not,  what is the difference ? How 
does this hypothetical vacuum energy density lose energy of the order of 
$10^{120}$ in transitioning from the Planck to the Quark-Hadron era ?

These and other difficulties \cite{LEE, ALIN} lead to the 
questioning of the usefulness of a hypothetical vacuum state.

In conclusion, it is clear that the concept of a vacuum state has lead to 
real confusion and difficulties in calculating ``real'' world 
physical quantities.
An analogous definition of a vacuum state will be proposed in 
the following sections.

\subsection{\Large{Origin of the "Effective Mass" - Gravitational Self Energy}}

Zel'dovich tried to relate a theory of elementary particles 
to the gravitational interaction 
energy associated with the Cosmological Constant. He 
argued intuitively that the vacuum energy density should be given by 
the gravitational interaction energy of virtual particles whose 
 self-energy was identically equal to zero.

The objective of the next section will be to 
relate the Cosmological Constant, a
macroscopic quantum mechanical parameter, with an effective mass, and a 
microscopic theory of nature, this can be shown  as: 
\begin{equation}
\Lambda_{Epoch\;Energy\;Scale}({\bf Macroscopic \;Parameter}) 
\leftrightarrow m^6_{EES}\leftrightarrow 
\epsilon_{Vac}^{EES} ({\bf Microscopic \;Theory})
\end{equation}
A speculative microscopic theory of gravity will be outlined in the 
following sections.

\subsection{Quantum Gravity}
\subsubsection{Theoretical Approach}

The approach will be to attempt to apply 
String theory to the Cosmological Constant problem.  Recently 
there have been advances in the non-perturbative regime of the theory, 
the most interesting developments have been in the fields of M
and D-brane theories.

M -Theory \cite{WITTEN} is a unique theory whose moduli 
space connects the five perturbative 
ten-dimensional String theories with 11-dimensional supergravity.
 It can be described in terms of 2 
ten-dimensional worlds where gauge and matter fields are localised, 
the connection 
between the two being a line segment or orbifold. In this 
model (see figure 3) gravity propagates in the 
 eleven-dimensional space of the bulk. 
\begin{figure}[h]
\epsfxsize=6.2in
\epsfysize=3.3in
\epsffile{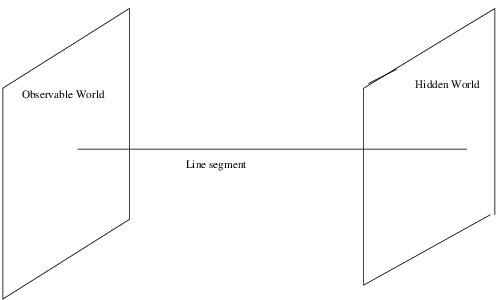}
\caption{M-theory picture of the world}
\end{figure}

The low energy limit of M-theory is 
equivalent to 11-D supergravity and 
corresponds to the strong coupling limit of
 a $E_8\times E_8$ 10-d heterotic String theory. 
On compactification, this gives rise to
 2 four-dimensional worlds separated by a line segment where gravity 
propagates in the bulk 5-d space-time. The theory predicts unification 
of gauge couplings at $10^{16}$GeV.

D-branes \cite{POLCHI} are defined as 
hypersurfaces (dynamical objects) on which 
open strings  end, here the D stands for  Dirichlet. 
The coordinates of the attached strings satisfy Dirichlet boundary conditions 
in the directions normal to the brane and Newmann conditions in the 
directions tangent to the brane.
A D-brane extending over p-flat spatial dimensions is 
described by the boundary conditions, given by,
\begin{equation}
\partial_{\perp} X^{\alpha=0,\cdots,p}=X^{m=p+1,\cdots,9}=0
\end{equation}

In the discussion of target-space duality in section 3.1.2 the 
spectrum of closed strings was shown to be invariant 
under the transformation $R\rightarrow \alpha^{'}/R$.  As 
$R\rightarrow 0$ the momentum states become 
massive but the winding modes approach a continuum (become very light), 
indicating the formation of a non-compact dimension.

D-branes appear in the 
$R\rightarrow 0$ limit of open string theory \cite{POLCHI}, where 
compactified on a small torus is equivalent to 
compactification on a large torus but with the open string end points 
restricted to lie on the hypersurfaces. D-branes are 
solitonic objects whose masses are given by 
$m_s/g_s,\; m_s=1/\sqrt{a^{'}}$, where 
$g_s$ is the String coupling constant.

Figure 4 shows 2 D-branes connected by an open string.
Figure 5 shows the interaction of two D-branes through the 
exchange of a closed String (graviton).

\begin{figure}[h]
\epsfxsize=6.2in
\epsfysize=3.3in
\epsffile{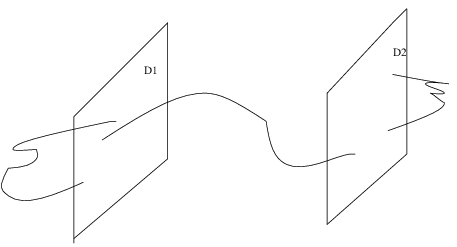}
\caption{Open strings streched between D-branes and with both ends on the 
same defect}
\end{figure}

\begin{figure}[h]
\epsfxsize=4.2in
\epsfysize=3.3in
\epsffile{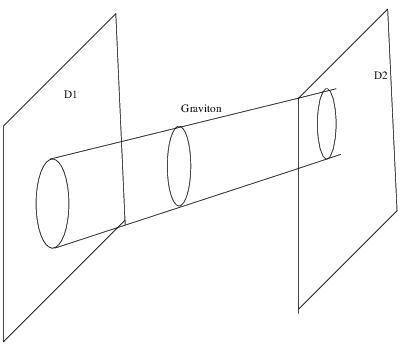}
\caption{Two D-branes interacting through the exchange of a closed string}
\end{figure}
When the separation between branes is of a comparable length to the 
Planck scale, the solitons can interact with 
each other via the exchange of open 
strings. It is this property of D-brane dynamics that makes them 
good probes for the small-scale structure of space-time and 
which may lead to a formal derivation of the 
second term in the modified Heisenberg Uncertainty Principle equation 
\cite{NICK}.
\begin{equation}
\Delta x \geq \frac{\hbar}{\Delta p}+\alpha^{'} \frac{\Delta p}{\hbar}.
\label{Uncer}
\end{equation}

\subsubsection{Vacuum State - Analogue Definition}

It was suggested previously that difficulties arise when trying to calculate 
physical quantities from a vacuum state. In this section we 
will associate the lowest energy or ground state of the Universe 
in terms of the product of the lowest energy states of 
the real and compactified worlds.

Here, in a leap of speculative imagination, 
the compactified world will be considered to act as a kind of vacuum, 
unseen but affecting the real world  through the 
interaction of gravity. The difference is that the hidden world 
is a physical entity, unlike the vacuum, which will allow the 
computation of physical quantities. Presently these calculations 
are not feasible due to technical difficulties, however 
there is no {\em {a priori}} reasons why they are not possible.

The main justification for the approach is taken from analogy 
with M-Theory where it is suggested
that our four-dimensional world is a brane embedded in a higher 
dimensional bulk space-time and  where all 
the matter fields (quarks, leptons,etc) together with the three gauge 
interactions 
live exclusively on the brane. Again, gravity is the only 
interaction that lives in the bulk space.

\subsubsection{Model - Approach for Calculating the Cosmological Constant}
Outlined below is a speculative 
approach for calculating the Cosmological Constant for the 
different epochs of the Universe. The Cosmological Constant will be  
associated  with the gravitation interaction energy - Casimir 
effect. This can be interpreted as the energy due to vacuum 
fluctuations of open strings stretched between two D-branes (
figures 4,5). The tree-level expression for the gravitational interaction  
is given by \cite{POLCHI}:
\begin{equation}
\epsilon(r,T)=-\frac{V_{(p)}}{2}\int \frac{d^{p+1} k}{(2 \pi)^{p+1}}
\int_0^{\infty} \frac{dt}{t} {\rm Str} e^{-\pi t(k^2+M(T)^2)/2},
\end{equation}

\begin{equation}
\epsilon(r,T)=-2\times \frac{V_(p)}{2}\int_{0}^{\infty}\frac{dt}{t}
(2\pi^2 t)^{-(p+1)/2} e^{{-r^2 t}{2\pi}}(-1/2)\sum_{s=2,3,4}(-)^s 
\frac{\theta_s^4(0|\frac{it}{2})}{\eta^{12}(\frac{it}{2})}F(T,\alpha^{'},
r,t)
\end{equation}
The second equation is derived from the first for the 
particular example shown in figure 5, the 
$\theta_i$'s are the Jacobi theta functions and $\eta(t)$
is the Dedekind eta function.

There is an expectation that the gravitational interaction 
energies for the various epoch (Planck, GUT, 
Electro-Weak, Quark-Hadron), namely the Cosmological Constants, will be 
recovered at some distance and temperature.

Interestingly, in the 
high temperature limit of string theory  the Hagedorn temperature 
leads 
to a phase transition which is analogous to the Quark-Hadron 
phase transition \cite{HAGEDORN}.

\subsubsection{Negative $\Lambda$ and Brown-Teitelboim mechanism}

The mechanism suggested by Brown and Teitelboim 
\cite{BROWN} has the appealing feature 
that a small negative Cosmological Constant compatible with experimental 
results can be reconciled with a large positive Cosmological Constant 
at an earlier period and thus with Inflation. In this mechanism, a 
Cosmological Constant is neutralized by nucleation of membranes associated
with antisymmetric fields.  
However, in order that their mechanism leads to a realistic value for the 
vacuum energy density the tension of the brane and its coupling constant 
are severely constrained.
Recently, Bousso and Polchinski use
\cite{BOUSSO} more than one four-forms
(the field strength of the three-form) to create a small $\Lambda_{eff}$.
As has been emphasized by the authors an associated 
difficulty with this approach is 
the stabilization of the compact dimensions 
\footnote{In principle extreme  regions of the string moduli 
can lead to small charges and brane tensions, however it is unclear 
how a moduli potential can be minimised in such regions.}, which is 
a main issue 
in string theory. 
The mechanism  of \cite{BROWN} combined with 
a string theory approach  
can in principle lead to a $negative$ effective Cosmological Constant 
compatible in magnitude and 
sign with that needed to explain the Galactic Velocity Rotation 
Curves.

\subsubsection{Summary}
In this part of the paper a speculative connection has been made 
between the 
vacuum and the hidden world of M-theory, and the 
Cosmological Constant and the gravitational interaction, 
energy given by String Theory. A suggestion has been made as 
how, at specific distances and temperatures, the different 
energy scales of the epochs could be recovered.

Note that in order to arrive at a non-zero value for the 
gravitational interaction   an assumption has made that there 
is soft supersymmetry-breaking  (the expression for the 
gravitational energy is identically equal to zero 
for exact supersymmetry) and higher-order effects are suppressed.

The application of String theory to the 
determination of cosmological parameters will be of 
considerable interest in the future.

\newpage

\section{Discussion}

This paper describes how the non-gravitational contribution to 
Galactic Velocity Rotation Curves can be explained in terms of a negative 
Cosmological Constant 
($\Lambda$).

It is shown that, within the Weak Field Approximation, 
the experimental values for the 
Cosmological Constant can be determined from galactic velocity 
rotation curves. A representative value has been determined to be of the 
order $|\Lambda_{Exp}|=2.0\times 
10^{-55} cm^{-2}$. This value compares
favourably with the theoretical value, derived from the Extended LNH, of 
$|\Lambda_{Theory}|=2.1\times 10^{-56} cm^{-2}$.

The Extended LNH was used to predict values for other cosmological 
parameters such as: Gravitational, Gravitational Modification Constants, and 
the effective mass density. The 
Weak Field Approximation was used to derive the modified Newton 
Equation, and String Theory via Duality, was used to establish 
the requirement for a fundamental length in nature.

The experimental results presented support the ideas of a long range 
fifth force, gravitational in nature, leading to a decelerating Universe, 
and that the Cosmological Constant is a good candidate for providing the 
missing "Dark Energy" of the Universe.




Matthews \cite{MATTHEWS} pointed out 
that, in the Extended Large Number Hypothesis,  replacing
the mass of the Proton by 
the effective mass of the vacuum energy density of the 
relevant epoch lead to agreement with experimental observation. 
The presently 
observed Universe, 
seen in terms of its cosmological 
 parameters, is well described  
by the Quark-Hadron energy scale.
However, the effective mass parameter has been introduced in an 
arbitrary way. The validity  of this term in determining  cosmological 
parameters will need to be justified  in terms of a more fundamental 
microscopic quantum theory.

A speculative theory for the evolution of the universe was outlined. 
It was shown that within the 
Extended LNH only two parameters are required to fully 
specify the cosmological parameters for that epoch: fundamental 
length and the vacuum energy density.

It was suggested that the concept of a vacuum leads to confusion and 
the inability to calculate ``real'' world quantities. A
speculative analogous 
physical alternative was proposed for the vacuum.


It was suggested that the application of String Theory to the 
problem of the 
Cosmological Constant 
could lead  to a theory of quantum gravity.

Finally, Dirac's belief in an as yet unknown 
fundamental theory linking the quantum 
mechanic origin of the Universe to 
large-scale cosmological parameters  may yet come true.

\section*{Acknowledgements}
We would like to thank A. Love, J. Hargreaves and 
D. Bailin for suggestions and 
useful discussions and  acknowledge the correspondence with R. Matthews.
We are grateful to  P. Salucci for useful discussions.
Finally we thank 
Mrs Deja Whitehouse for proof-reading the paper.
GVK was supported by PPARC.

\end{document}